\documentclass[aps,english,prb,reprint,superscriptaddress, 
citeautoscript,showpacs]{revtex4-1}
\usepackage{graphicx}
\usepackage{amssymb}
\usepackage{babel}
\usepackage{color}
\usepackage{multirow}
\usepackage{setspace}
\usepackage{longtable}
\begin{document}

\title{Biphenylene monolayer as a two-dimensional nonbenzenoid carbon allotrope: A first-principles study}

\author{A. Bafekry}\email{bafekry.asad@gmail.com}
\affiliation{Department of Radiation Application, Shahid Beheshti University, Tehran 1983969411, Iran}
\author{M. Faraji}
\affiliation{Micro and Nanotechnology Graduate Program, TOBB University of Economics and Technology, Sogutozu Caddesi No 43 Sogutozu, 06560, Ankara, Turkey}
\author{M. M. Fadlallah}
\affiliation{Department of Physics, Faculty of Science, Benha University, 13518 Benha, Egypt}
\author{H. R. Jappor}
\affiliation{Department of Physics, College of Education for Pure Sciences, University of Babylon, Hilla, Iraq}
\author{S. Karbasizadeh}
\affiliation{Department of Physics, Isfahan University of Technology, Isfahan, 84156-83111, Iran}
\author{M. Ghergherehchi}
\affiliation{Department of Electrical and Computer Engineering, Sungkyunkwan University, 16419 Suwon, Korea}
\author{D. Gogova} 
\affiliation{Department of Physics, University of Oslo, P.O. Box 1048, Blindern, Oslo, Norway}

\begin{abstract}
In a very recent accomplishment, the two-dimensional form of Biphenylene network (BPN) has been successfully fabricated [Fan et al., Science, 372, 852-856 (2021)]. Motivated by this exciting experimental result on 2D layered BPN structure, herein we perform detailed density functional theory-based first-principles calculations, for the first time, in order to gain insight into the structural, electronic and optical properties of this promising nanomaterial. Our theoretical results reveal the BPN structure is constructed from three rings of tetragon, hexagon and octagon, meanwhile the electron localization function shows very strong bonds between the C atoms in the structure.
The dynamical stability of BPN is verified via the phonon band dispersion calculations. The mechanical properties reveal the brittle behavior of BPN monolayer. The Young’s modulus has been computed as 0.1 TPa, which is smaller than the corresponding value of graphene, while the Poisson’s ratio determined to be 0.26 is larger than that of graphene. The band structure is evaluated to show the electronic features of the material; determining the BPN monolayer as metallic with a band gap of zero. The optical properties (real and imaginary parts of the dielectric function, and the absorption spectrum) uncover BPN as an insulator along the zz direction, while owning metallic properties in xx and yy directions. We anticipate that our discoveries will pave the way to the successful implementation of this new 2D allotrope of carbon in advanced nanoelectronics.
\end{abstract}

\maketitle

\section{Introduction}
Carbon can be found in a variety of allotropes. Aside from diamond and graphite, other newly found types with remarkable characteristics have been identified due to the carbon tendency to be present in several hybridizations. Carbon's ability to form in a variety of shapes through different dimensions has sparked interest in discovering additional allotropic forms made up of diverse networks of rings and polygons. To clarify, researchers have identified a wide range of carbon allotropes composed of sp$^3$, sp$^2$, and sp hybridized carbon atoms such as graphenylene \cite{1}, graphdiyne \cite{2}, graphyne \cite{3}, cyclocarbons \cite{4}, pentaheptite \cite{5}, graphene \cite{6}, phagraphene \cite{7}, haeckelites \cite{8}, pentahexoctite \cite{9}, carbon nanotubes \cite{10}, carbon nanocone \cite{11}, and fullerene \cite{12}. Owing to their electronic and mechanical properties, these materials have been widely advocated as a possible candidate for popularly used applications such as sensors \cite{13}, energy storage \cite{14}, nanoelectronic devices \cite{9}, thermal rectifier \cite{15}, etc. Among these forms of carbon allotropes, two-dimensional (2D) graphene and its like have been extensively studied theoretically and experimentally \cite{16,17,18,19,20}. In addition, graphene has piqued the curiosity of scientists in a wide range of domains. To be more specific, this 2D material is a single, thin sheet of carbon atoms organized in a hexagonal honeycomb pattern that is one atom thick. The immense attention for graphene is due to the specific characteristics that set it apart from its 2D brethren. To exemplify, graphene's most impressive features are thermal, electronic, and mechanical properties such as its large surface area \cite{21}, brilliant high Young's modulus \cite{22}, excellent thermal conductivity \cite{23}, and electronic mobility \cite{24}.

Although recent technological advances have led to tremendous progress in the creation and manufacture of many carbon allotropes, there are forms in which carbon atoms themselves can be organized that have still to be discovered. This was confirmed by a very recent experimental study carried out by scholars who identified a novel kind of non-benzene carbon named biphenylene which is as thin as graphene and thus as thin as an atom \cite{25}. This novel carbon grid is made up of octagons, hexagons, and squares of sp$^2$ carbon atoms that were derived from a regular grid. According to the researchers, this material appeared to have fascinating electronic characteristics that differ from those of graphene. They synthesized this non-benzene carbon by bottom-up approaches and used high-resolution scanning probe microscopy to establish the network's unique structure and discovered that its electronic characteristics are significantly distinct from that of graphene. Importantly, authors found that the measured bandgap decreased with increasing the widths of biphenylene ribbon. Despite this, the new form of carbon boasts metallic properties thanks to its narrow webbing stripes that are 21 atoms wide. However, a non-benzene 2D carbon polymorph containing rings that have 6 or more atoms will possibly give better anode materials in lithium-ion rechargeable batteries because of its expected higher lithium storage capacity than graphene \cite{26}. Precisely, these stripes can be utilized as conductive wires in lithium-ion batteries and carbon-based electronic systems.

In this paper, motivated by the very recent successful experiments on the biphenylene, we investigate the structural and electronic properties of named biphenylene network using first-principles simulations. The present discoveries constitute a significant breakthrough in our knowledge of the properties of these nanostructures. We anticipate that our discoveries will pave the way to the successful implementation of this new 2D allotrope of carbon in future nanoelectronics and optoelectronic devices.

\section{Method}
The density-functional theory (DFT) calculations in this work are performed using the plane-wave basis projector augmented wave (PAW) method along with generalized gradient approximation (GGA) with Perdew-Burke-Ernzerhof(PBE)\cite{GGA-PBE1,GGA-PBE2} functional as implemented in the Vienna \textit{ab-initio} Simulation Package (VASP)\cite{vasp1,vasp2}. Moreover, for the band structure calculations Heyd-Scuseria-Ernzerhof (HSE06)\cite{hse} screened-nonlocal-exchange functional of the generalized Kohn-Sham scheme is included for more accurate band gap calculations. The kinetic energy cut-off of 500 eV is set for plane-wave expansion and the energy is minimized for the structure until variation in the energy falls below 10$^{-8}$ eV. To get optimized structures, total Hellmann-Feynman forces are reduced to 10$^{-7}$ eV/\AA {}. 
The k-points for sampling over the Brillouin zone (BZ) integration are generated using the Monkhorst-Pack scheme \cite{Monkhorst} with 21$\times$21$\times$1 $\Gamma$ centered \textit{k}-point sampling. Charge transfer analysis is accomplished using the Bader technique\cite{Henkelman}. The empirical dispersion method of DFT-D3 \cite{DFT-D3} is employed to get insight into the van der Waals interactions. The vibrational properties are obtained from the small displacement method as implemented in the PHONOPY code \cite{phon}. Simulated scanning tunneling microscopy (STM) images are obtained using the Tersoff-Hamann theory \cite{Tersoff} and are graphed using WSxM software \cite{WSxM}. The training set is prepared by conducting ab-initio molecular dynamics (AIMD) simulations over $2\times2\times1$ supercells with $2\times2\times1$ k-point grids. 

\section{Structural Properties}

\begin{figure}[!b]
\includegraphics[scale=1]{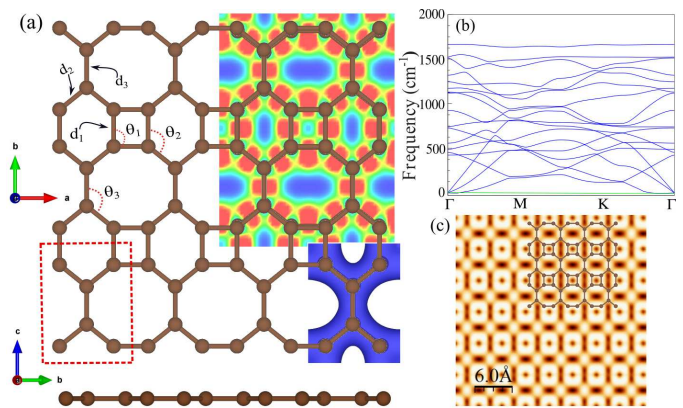}
\caption{ (a) Atomic structure and electron localization function (ELF); (b) phonon band dispersion, and (c) STM simulated image of the BPN monolayer. Square primitive unit cell is indicated by the red dotted line. The electron localization function (ELF) and total charge density are indicated as inset.}
\label{1}
\end{figure}

\begin{table*}[!htb]
	\centering
	\caption{\label{table1} 
		Structural and electronic parameters of the optimized BPN monolayer as shown in Fig. 1(a), including: 
		lattice constants \textit{a} and \textit{b}, 
		bond lengths of C-C atoms in tetragon ring ($d_{1}$), hexagon ring ($d_{2}$) and octagon ring ($d_{3}$), 
		bond angles of C-C-C atoms in tetragon ring ($\theta_{1}$), hexagon ring ($\theta_{2}$) and octagon ring ($\theta_{3}$), 
		cohesive energy per atom, $(E_{coh})$ and 
		electronic states $(ES)$ which are specified as metal (M).
		Work function $\Phi$; shear modulus (S); Young's modulus (Y); and Poisson's ratio ($\nu)$) are also inserted into the table.
	}
	\scalebox{1}{ 
		\begin{tabular}{ccccccccccc}
			\hline
			Sys. &\textit{a} &\textit{b} & \textit{d$_{1,2,3}$}& \textit{$\theta_{1,2,3}$}& $E_{coh}$ & $ES$ & $\Phi$ & S & Y& $\nu$ \\ 
			& (\AA) & (\AA) & (\AA) & ($^{\circ}$) & (eV/atom) & &(eV)& (TPa) &(TPa)& \\
			\hline
			BPN  & 3.75  & 4.52  & 1.45,1.40,1.44  & 90,109,125 & -7.55 & M & 4.30 & 0.03 & 0.10 &  0.26\\
			\hline
		\end{tabular}}
	\end{table*}
	
The atomic structure of the BPN monolayer in the top and side view is illustrated in Fig. \ref{1}(a). 
BPN possesses a carbon lattice with a rectangular primitive unit cell (indicated as a red rectangle) formed by six C atoms. 
The lattice constants of BPN are calculated to be 3.75 \AA{} (\textit{a}) and 4.52 \AA{} (\textit{b}). In the BPN structure, we can see three types of C-C rings of tetragon, hexagon and octagon. The bond lengths in the C-C tetragon ring is determined to be $d_{1}$=1.45 \AA{} and the bond angle is 90$^{\circ}$. In the C-C hexagon ring, $d_{1}$ is equal to 1.40 \AA{} and the bond angle of C-C-C is 109$^{\circ}$, while in the C-C octagon ring $d_{3}$ is 1.44 \AA{}, with the respective bond angle of the C-C-C being 125$^{\circ}$. The optimized structure of the BPN monolayer has a planar lattice and no C atom deviates from this two dimensional sheet. The structural parameters are gathered and given in Table \ref{table1}. 
To get a better understanding of the underlying characteristics of bonding properties, the electron localization function (ELF) along the (0 0 1) plane and total charge density are quantified and illustrated as insets in Fig. \ref{1}(a). 
This sum takes a value between 0 and 0.5, where ELF = 0.5 corresponds to perfect localization. The ELF value between C atoms is around the maximum amount of 0.50, which shows a large proportion of bonds in the BPN monolayer as very strong covalent bonds. This is clearly depicted in Fig. \ref{1}(a), where the color red shows intense localization and in turn, a stronger bond. The dynamical stability of BPN is verified using phonon band dispersions shown in Fig. \ref{1}(b). As seen in this figure, the phonon band structure does not have any of the bands enter the negative realm and is therefore, stable. 
The STM simulated image is obtained from first-principles DFT calculations in order to provide visible guidance for future experimental observations (see Fig. \ref{1}(c)). The atomistic structure is easily recognized from the predicted STM image, where the C atoms are brighter. 

The cohesive energy per atom that quantifies the stability of a material is calculated using the following equation:
\begin{equation}
E_{coh} = [E_{tot}-n_{C}E_{C}]/(n_{C}),
\end{equation}
where $E_{C}$ represents the energy of an isolated single C atom, $E_{tot}$ represents the total energy of the BPN and n$_{C}$ stands for the number of C atoms in the primitive unit cell. The negative cohesive energy for this layer suggests that the free-standing state of the structure is likely to be stable, which is a very important clue for technologists in their quest for new planar sp$^2$
-hybridized carbon allotropes. The calculated cohesive energy for BPN is -7.55 eV/atom, and is shown in Table \ref{table1}.
It is found out that the electrostatic potential of the BPN monolayer is flat in the vacuum region. The work function of BPN is calculated using the following formula: 
$\Phi =E_{vacuum}-E_{F}$ 
and is determined to be 4.30 eV.

The mechanical properties of the BPN monolayer evaluated using harmonic approximation indicate that all elastic constant values (13 independent elastic constants) satisfy the Born's criteria, which confirms the mechanical stability of BPN. Young's modulus (0.10 TPa) is smaller than the experimental (2.4 TPa) \cite{ex} and theoretical (1.0 TP) values of graphene \cite{A31,A32}. We find that shear modulus is 0.03 TP and Poisson's ratio is 0.26, which is larger than that of graphene (0.16) \cite{QS} and is very close to that of MoSi$_{2}$N$_{4}$ (0.28) \cite{our}. The value of Poisson's ratio indicates that the BPN monolayer has a brittle behavior (it is less than 0.33 \cite{A33}).

\section{Electronic Properties}

\begin{figure}[!b]
\centering
\includegraphics[scale=1]{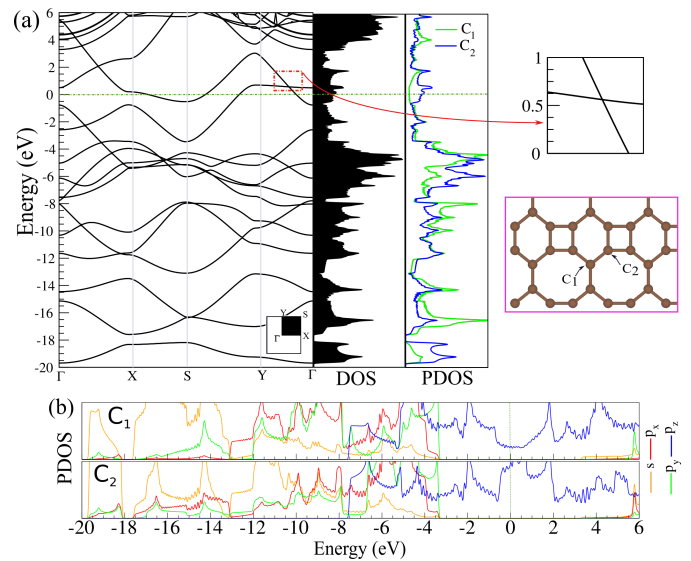}
\caption{ (a) Electronic band structure and (b) DOS/PDOS of the BPN monolayer. (b) Calculated PDOS of atoms with respect to $s$, $p_{x,y}$ and $p_z$ orbitals states.}
\label{2}
\end{figure}

The results of the electronic structure calculations, i.e., the band structure, density of states (DOS) and partial DOS (PDOS) of the BPN monolayer are presented in Fig. \ref{2}(a,b). In the bands portrayed here, no space is seen unaltered between the VBM and the CBM, resulting in mettalic features in the strucutre. The dirac cone is formed slightly above the Fermi level and is severely tilted to the side. This tilt has been observed in other Dirac/Weyl materials \cite{Katayama2006,Kajita2014, Zhou2014} and the solid-state system containing this tilted cone have been interpreted as systems where the effective spacetime is non-Minkowski and it is, in turn, deformed \cite{Mola2019}. This indeed multiplies the importance of the BPN monolayer and opens a new path for further investigation. Total DOS of the system is also displayed in Fig. \ref{2}(a), with the addition of a DOS where the effects from C$_1$ and C$_2$ atoms are seen. These atoms represent the only symmetries where the DOS is specific and all the other atoms of the system take the same shape as one of the chosen particles. This can be understood given the mirror symmetries present in the two dimensional sheet. The noteworthy part of this DOS is near the Fermi level, where the atom of C$_2$ is in full effect, contrary to C$_1$ which gives a very little contribution. The position of these two atoms is illustrated in Fig. \ref{2}(a). PDOS of the indicated atoms is drawn in Fig. \ref{2}(b). In this shape, the $s$, $p_x$, $p_y$ and $p_z$ shells are given separately. PDOS of C$_1$ and C$_2$ show striking resemblance in some sections, but are wildly different in other areas. Both atoms show a heavy involvement of $p_z$ shell near the Fermi level; while, as mentioned before, only C$_2$ is intensely contributing when it gets close to the line. It is also witnessed for both atoms that with getting further away from the Fermi level, the $p_z$ shell subsides completely and abruptly. This happens as the $p_x$ shell also ceases to exist almost completely in the atom of C$_1$ after going deep in the states, while the same thing does not happen in C$_2$ and $p_x$ continues to contribute strongly even further down the line.  

\section{Optical properties}

\begin{figure}[!t]
	\includegraphics[scale=1]{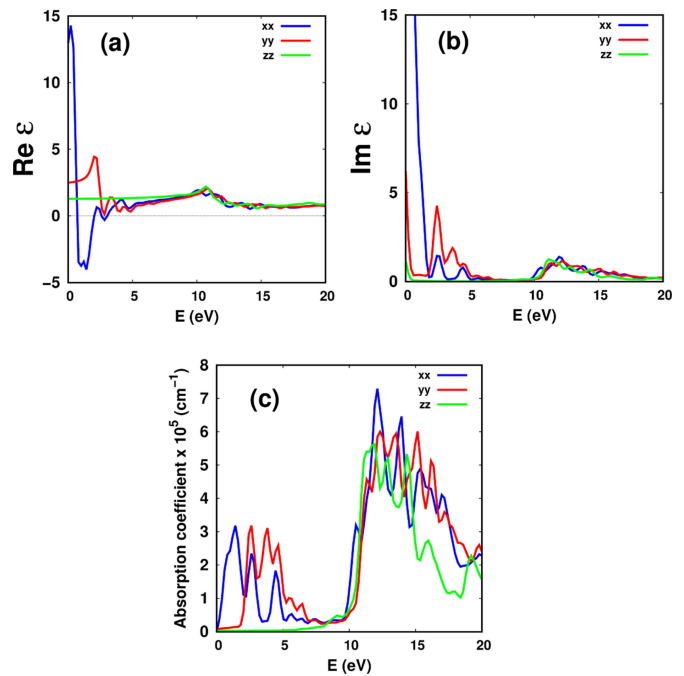}
	\caption{(a) Real (Re($\varepsilon$)), and (b) Imaginary (im($\varepsilon$)) parts of the dielectric function, and (c) the absorption coefficient vs. the photon incident energy.}
	\label{3}
\end{figure}

The real and imaginary parts of the dielectric function of BPN monolayer are illustrated in Figs. \ref{3}(a),(b). These are typically used to describe the monlolayer response to the electromagnetic field. That is, wherever a part of the dielectric function is negative, electromagnetic waves become weak and absorption or reflection occurs. The real and imaginary parts of the dielectric function are connected by the Kramers-Kronig relations. The dielectric function has two interband and intraband transitions; the interband transition is due to the excitation of the absorption edges, however, the intraband transition happens because of the volume of plasmon excitation. The imaginary part of the dielectric function is calculated by taking into account all possible transitions from occupied to unoccupied states. In Fig. \ref{3}(a), the static value of the real part has shown that its value is 12.5, 2.5 and 1.3 for xx, yy and zz components, respectively. Many peaks appear in the xx and yy directions from 0.5 eV to 5.0 eV. Furthermore, many small peaks appear in all directions when moving past 10.0 eV. According to the imaginary part of the dielectric function (Fig. \ref{3}(b)), it can be seen that the value of the imaginary part in xx direction tends to be infinitely positive, which represents the extremely intense metallic behavior. Similar to the real part, the imaginary part also owns many peaks above 10.0 eV in all directions. One can notice that the value of the imaginary part is zero from 5.0 eV to 10 eV in yy direction and from 0.5 to 10.0 eV in zz direction. Optical absorption depends on the imaginary part of the dielectric tensor, where the electrons in valence band states are induced by absorbing a photon into the conduction band states. The absorption peaks in the yy direction, from 0 eV to 5 eV, are shifted towards larger energies, which are related to the behavior of real and imaginary dielectric constants. In addition, the zero absorption in the zz direction is related to the zero value of the imaginary part and a constant value of the real part within the same energy range from 0.0 eV to 10 eV. The absorption amount is very large at energies higer than 10.0 eV for all components (Fig. \ref{3}(c)). The xx and yy components have absorption peaks in the energy range from 0.0 eV to 7.5 eV, meanwhile there is no any contribution in the zz direction. This phenomenon dictates metallic properties in xx and yy directions and insulating behavior along zz direction.

\section{Conclusion}
The structural, electronic and optical properties of the BPN monolayer are investigated by means of first-principles calculations. 
The overall shape of the BPN structure is constructed from three rings of tetragon, hexagon and octagon, meanwhile the electron localization function shows very strong bonds between the C atoms in the structure.
The phonon band dispersion provides a substantial proof for the dynamical stability of BPN, as no bands have values of less than zero. The STM simulated image has been evaluated and shows a picture that can be extremely useful in order to observe the structure in future experimental endeavors. 
BPN gives a cohesive energy of -7.55 eV/atom, while the work function value is determined to be 4.30 eV. 
Mechanical properties calculations have revealed the brittle behavior of the BPN monolayer. 
The electronic band structure reveals the BPN monolayer as a metal with conduction and valence bands overlapping considerably on the Fermi energy level. A tilted Dirac cone does appear in the bands, slightly higher than the Fermi energy level. 
The optical properties, are calculated giving valuable information about where absorption or reflection is bound to happen. 
These two parts of the dielectric function, along with the absorption coefficient vs. the photon incident energy, demonstrate the metallic properties of the BPN in xx and yy direction and insulating behavior along the zz direction.

\section{Conflicts of interest}
The authors declare that there are no conflicts of interest regarding the publication of this paper.

\section{ACKNOWLEDGMENTS}
This work was supported by the National Research Foundation of Korea (NRF) grant funded by the Korea government (MSIT) (NRF-2017R1A2B2011989).

\end{document}